\documentclass[aps,showpacs]{revtex4}
\usepackage{graphicx}
\begin{document}

\title{Thermal freeze-out versus chemical freeze-out reexamined}
\author{Dariusz Prorok}
\affiliation{Institute of Theoretical Physics, University of
Wroc{\l}aw,\\ Pl.Maksa Borna 9, 50-204  Wroc{\l}aw, Poland}
\date{August 19, 2009}

\begin{abstract} An alternative, to the commonly used blast-wave, model describing
the freeze-out hypersurface is applied to fit the $p_{T}$-spectra of
identified hadrons measured at relativistic heavy-ion collisions at
$\sqrt{s_{NN}}=62.4,\; 130$ and 200 GeV. Decays of resonances are
taken into account completely. It has turned out that the fitted
freeze-out temperature and baryon number chemical potential depend
weakly on the centrality of the collision and their values are close
to the chemical freeze-out values determined from fits to particle
yield ratios.
\end{abstract}

 \maketitle

\section {Introduction}
\label{Intrd}

During a heavy-ion collision a hot and dense medium is created which
eventually evolves into a state of freely streaming particles. The
process of hadron decoupling is called \textit{freeze-out} and two
kinds of freeze-out are distinguished
\cite{Heinz:1999kb,Heinz:2006ur}: (i) \textit{chemical freeze-out}
at $T_{chem}$ when the hadron abundances become fixed and (ii)
\textit{thermal (kinetic) freeze-out} at $T_{kin}$ when elastic
rescattering processes cease and hadrons start to escape freely. And
$T_{chem} \geq T_{kin}$ is expected. Values of the statistical
parameters at the chemical freeze-out are determined from fits to
particle yield ratios, whereas corresponding values at the kinetic
freeze-out are fitted to the spectra of hadrons. From here forward,
statistical parameters at the kinetic freeze-out mean their values
fitted to the spectra. $T_{chem} \sim 150-170$ MeV is estimated at
highest heavy-ion reaction energy
\cite{Torrieri:2004xu,Kaneta:2004zr,Florkowski:2001fp,Cleymans:2004pp,Braun-Munzinger:2001ip,Rafelski:2004dp,Adams:2003xp,Barannikova:2004rp,Barannikova:2005rw,Baran:2003nm,Braun-Munzinger:2003zd}.
Additionally, fits done for various centrality classes have revealed
that $T_{chem}$ is almost independent of centrality
\cite{Kaneta:2004zr,Cleymans:2004pp,Rafelski:2004dp,Adams:2003xp,Barannikova:2004rp}.
On the contrary, the temperature at the kinetic freeze-out depends
on the centrality and is substantially lower. From the most central
to the peripheral bin it changes as follows: $T_{kin}=121-161$ MeV
for PHENIX at $\sqrt{s_{NN}}=130$ GeV \cite{Adcox:2003nr},
$T_{kin}=111-147$ MeV for PHENIX at $\sqrt{s_{NN}}=200$ GeV
\cite{Burward-Hoy:2002xu}, $T_{kin}=89-129$ MeV for STAR at
$\sqrt{s_{NN}}=200$ GeV \cite{Adams:2003xp} and $T_{kin}=110-115$
MeV for BRAHMS at $\sqrt{s_{NN}}=200$ GeV \cite{Arsene:2005mr}. For
the PHOBOS data at $\sqrt{s_{NN}}=62.4$ GeV $T_{kin}=103, 102, 101$
MeV for the central, mid-peripheral and peripheral bin, respectively
\cite{Back:2006tt}. However, the aforementioned estimates of
$T_{kin}$ have been done within the very simplified hydrodynamic
model, \textit{i.e.} the blast-wave model
\cite{Schnedermann:1993ws}.

The main difference between fitting the statistical parameters at
the chemical freeze-out and at the thermal freeze-out is that the
first procedure is independent of a pattern of expansion. Thus,
assuming boost invariance and that statistical parameters are
constant on a freeze-out hypersurface one has

\begin{equation}
{(dN_{i}/dy)_{y=0} \over (dN_{j}/dy)_{y=0} } = {N_{i}  \over N_{j}}
= { n_{i} \over n_{j} }\;, \label{Ratiodens}
\end{equation}

\noindent where the last equality follows from the factorization of
the volume of the system (for more details see
\cite{Broniowski:2002nf,Cleymans:1998yf,Cleymans:1999st}) and the
density of particle species $i$ reads

\begin{equation}
n_{i}(T, \mu_{B}) = n_{i}^{primordial}(T, \mu_{B}) + \sum_{a}
\varrho(i,a)\; n_{a}^{primordial}(T, \mu_{B}) \;, \label{nchj}
\end{equation}

\noindent where $n_{a}^{primordial}(T, \mu_{B})$ is the thermal
density of the $a$th particle species at the freeze-out,
$\varrho(i,a)$ is the final number of particles of species $i$ which
can be received from all possible simple or sequential decays of
particle $a$ and the sum is over all species of resonances included
in the hadron gas. This means that values of the statistical
parameters at the chemical freeze-out do not depend on the assumed
model of expansion. Therefore fitting measured ratios can not
distinguish between the blast wave model and the present model. The
only sources of valuable information about the pattern of expansion
are measured spectra.

In this paper, it will be shown that the behavior of $T_{kin}$ is
model dependent and within a different hypersurface and with
complete treatment of resonance decays different conclusions about
statistical parameters at the kinetic freeze-out can be obtained.
Namely, the statistical parameters at the kinetic freeze-out are
roughly centrality independent and their values are close to the
corresponding values at the chemical freeze-out.

\section { Foundations of the model }
\label{Foundat}

The model applied here is inspired by the single-freeze-out model of
Refs.~\cite{Broniowski:2001we,Broniowski:2001uk}, but its main
assumption about the simultaneous occurrence of chemical and thermal
freeze-outs \textit{is not postulated}. The assumptions of the
present model are as follows. A noninteracting gas of stable hadrons
and resonances at chemical and thermal equilibrium appears at the
latter stages of a heavy-ion collision. The gas cools and expands,
and after reaching the freeze-out point it ceases. The conditions
for the freeze-out are expressed by values of two independent
thermal parameters: $T$ and $\mu_{B}$. The strangeness chemical
potential $\mu_{S}$ is determined from the requirement that the
overall strangeness equals zero. All confirmed resonances up to a
mass of $2$ GeV from the Particle Data Tables \cite{Hagiwara:fs},
together with stable hadrons, are constituents of the gas. The
freeze-out hypersurface is defined by the equation

\begin{equation}
\tau = \sqrt{t^{2}-r_{x}^{2}-r_{y}^{2}-r_{z}^{2}}= const \;.
\label{Hypsur}
\end{equation}

\noindent with the additional condition

\begin{equation}
r= \sqrt{r_{x}^{2}+r_{y}^{2}} < \rho_{max}. \label{Transsiz}
\end{equation}

\noindent which makes the transverse size of the system finite.
$\tau$ and $\rho_{max}$ constitute two geometric parameters of the
model.

The four-velocity of an element of the freeze-out hypersurface is
proportional to its coordinate

\begin{equation}
u^{\mu}={ {x^{\mu}} \over \tau}= {t \over \tau}\; \left(1,{ {r_{x}}
\over t},{{r_{y}} \over t},{{r_{z}} \over t}\right) \;.
\label{Velochyp}
\end{equation}

\noindent The shape of the hypersurface defined by
eqs.~(\ref{Hypsur})-(\ref{Transsiz}) and the above form of the flow
imply that the model is longitudinally boost-invariant and
cylindrically symmetric. The last feature is fulfilled strictly only
in the case of a central collision, but experimental spectra are
averaged over the azimuthal angle, which means that they look as if
they were uniform in this angle. In the view of
eq.~(\ref{Velochyp}), $\tau$ given by eq.~(\ref{Hypsur}) is the
usual proper time of the fluid element. Thus in this model the
freeze-out takes place at the same moment of time in the local rest
frame of a fluid element. In the c.m.s. of the colliding nuclei this
means that the farther particles are from the center of the
collision the later they freeze out. For the 1+1 dimensional Bjorken
model \cite{Bjorken:1983qr} this is the direct conclusion from the
form of its solution: all thermodynamical quantities depend only on
a local proper time $\tau = \sqrt{t^{2}-z^{2}}$, so if one assumes
that the freeze-out takes place at a given temperature, $T_{f.o.}$,
it means that it happens at one moment of the proper time. The same
is true for a spherically symmetric expansion in 1+3 dimensions at
sufficiently long times \cite{Cooper:1974qi}. Note that in that case
the freeze-out hypersurface has also the shape given by
eq.~(\ref{Hypsur}). The form of the four-velocity, as implied by
eq.~(\ref{Velochyp}), determines the three-velocity, $\vec{v}=
\vec{r}/t$. This is the so-called \textit{scaling solution}, which
is the exact solution for the Bjorken model \cite{Bjorken:1983qr}.
In 1+3 dimensions it was shown that for a spherically symmetric
expansion the scaling solution might develop if the sound velocity
squared satisfies $c_{s}^{2} < 0.2$ \cite{Cooper:1974qi}. For
boost-invariant, cylindrically symmetric systems recent results of
ref.~\cite{Chojnacki:2004ec} indicate that the scaling solution can
develop in 130 and 200 GeV Au-Au collisions within times 7-15 fm.
This is roughly the scale of the freeze-out initializing time
predicted in this analysis for central and mid-central bins (see
Table~\ref{Table1}).

The following parameterization of the hypersurface is chosen:

\begin{equation}
t= \tau \cosh{\alpha_{\parallel}}\cosh{\alpha_{\perp}},\;\;\; r_{x}=
\tau \sinh{\alpha_{\perp}}\cos{\phi},\;\;\; r_{y}=  \tau
\sinh{\alpha_{\perp}}\sin{\phi},\;\;\;r_{z}=\tau
\sinh{\alpha_{\parallel}}\cosh{\alpha_{\perp}}, \label{Parahyp}
\end{equation}

\noindent where $\alpha_{\parallel}$ is the rapidity of the element,
$\alpha_{\parallel}= \tanh^{-1}(r_{z}/t)$, and $\alpha_{\perp}$
controls the transverse radius $r=\tau \sinh{\alpha_{\perp}}$.

The maximum transverse-flow parameter (or the surface velocity) at
the central slice is given by

\begin{equation}
\beta_{\perp}^{max}= { \rho_{max} \over
{\sqrt{\tau^{2}+\rho_{max}^{2}}}}= { {\rho_{max}/\tau} \over
{\sqrt{1+(\rho_{max}/\tau)^{2}}}}\;. \label{Betmax}
\end{equation}

The invariant distribution of the measured particles of species $i$
has the form \cite{Broniowski:2001we,Broniowski:2001uk}

\begin{equation}
{ {dN_{i}} \over {d^{2}p_{T}\;dy} }=\int
p^{\mu}d\sigma_{\mu}\;f_{i}(p \cdot u) \;, \label{Cooper}
\end{equation}

\noindent where $d\sigma_{\mu}$ is the normal vector on a freeze-out
hypersurface, $p \cdot u = p^{\mu}u_{\mu}$ , $u_{\mu}$ is the
four-velocity of a fluid element and $f_{i}$ is the final momentum
distribution of the particle in question. The final distribution
means here that $f_{i}$ is the sum of primordial and simple and
sequential decay contributions to the particle distribution. The
primordial part of $f_{i}$ is given by a Bose-Einstein or a
Fermi-Dirac distribution at the freeze-out. A decay contribution is
a one-dimensional or multidimensional integral of the momentum
distribution of a decaying resonance (the exact formulae are
obtained from the elementary kinematics of a many-body decay or the
superposition of such decays, for details see
\cite{Broniowski:2002nf} and the Appendix in \cite{Prorok:2004af}).
The resonance is a constituent of the hadron gas and its
distribution is also given by the Bose-Einstein (Fermi-Dirac)
distribution function. Therefore the final distribution $f_{i}$
depends explicitly on $T$ and $\mu_{B}$.

With the use of eqs.~(\ref{Velochyp}) and (\ref{Parahyp}), the
invariant distribution (\ref{Cooper}) takes the following form:

\begin{equation}
{ {dN_{i}} \over {d^{2}p_{T}\;dy} }= \tau^{3}\;
\int\limits_{-\infty}^{+\infty}
d\alpha_{\parallel}\;\int\limits_{0}^{\rho_{max}/\tau}\;\sinh{\alpha_{\perp}}
d(\sinh{\alpha_{\perp}})\; \int\limits_{0}^{2\pi} d\xi\;p \cdot u \;
f_{i}(p \cdot u) \;, \label{Cooper2}
\end{equation}

\noindent where

\begin{equation}
p \cdot u = m_{T}\cosh{(\alpha_{\parallel}-y)}\cosh{\alpha_{\perp}}-
p_{T}\cos{\xi}\sinh{\alpha_{\perp}}\;. \label{Peu}
\end{equation}

The model has four parameters, the two thermal parameters, the
temperature $T$ and the baryon number chemical potential $\mu_{B}$,
and the two geometric parameters, $\tau$ and $\rho_{max}$. It should
be stressed that now all parameters of the model are fitted
simultaneously, opposite to the case of
Refs.~\cite{Broniowski:2001we,Broniowski:2001uk} where the
determination proceeded in two steps. First, statistical parameters
$T$ and $\mu_{B}$ were fitted with the use of the experimental
ratios of hadron multiplicities at midrapidity. Then geometric
parameters were determined from fits to the transverse-momentum
spectra. Therefore the assumption that the chemical freeze-out
happens simultaneously with the kinetic freeze-out (the single
freeze-out) was crucial in that approach. Now all parameters are
fitted to the spectra, so the aforementioned assumption is not
necessary and values of statistical parameters have the meaning of
the values at the kinetic freeze-out.

With the use of eq.~(\ref{Cooper2}) the measured transverse-momentum
spectra of $\pi^{\pm}$, $K^{\pm}$, $p$ and $\bar{p}$
\cite{Adams:2003xp,Adcox:2003nr,Adler:2003cb,Arsene:2005mr,Back:2006tt}
can be fitted to determine values of the parameters of the model
(data points with $p_{T} > 3$ GeV have been excluded). Fits are
performed with the help of the $\chi^{2}$ method. For the $k$th
measured quantity $R_{k}^{exp}$ and its theoretical equivalent
$R_{k}^{th}(\alpha_{1},...,\alpha_{l})$, which depends on $l$
parameters $\alpha_{1},...,\alpha_{l}$, the $\chi^{2}$ function is
defined as

\begin{equation}
\chi^{2}(\alpha_{1},...,\alpha_{l}) = \sum_{k=1}^{n} {
(R_{k}^{exp}-R_{k}^{th}(\alpha_{1},...,\alpha_{l}))^{2} \over
\sigma_{k}^{2}} \;, \label{Chidef}
\end{equation}

\noindent where $\sigma_{k}$ is the error of the $k$th measurement
and $n$ is the total number of data points. The fitted (optimum)
values of parameters mean the values at which $\chi^{2}$ has a
minimum.

\section {Results}
\label{Finl}

\begin{table*}
\caption{\label{Table1} Values of the statistical and geometric
parameters of the model for various centrality bins fitted with the
use of the RHIC final data for the $p_{T}$ spectra of identified
charged hadrons
\protect\cite{Adams:2003xp,Adcox:2003nr,Adler:2003cb,Arsene:2005mr}.
All data are at midrapidity, except the PHOBOS case (first three
rows) where data are at $y=0.8$ \protect\cite{Back:2006tt}. }
\begin{ruledtabular}
\begin{tabular}{c|ccccccccc} \hline Au-Au collision & Centrality & $N_{part}$ & $T_{kin}$ &
$\mu_{B}$ & $\rho_{max}$ & $\tau$ & $\beta_{\perp}^{max}$ &
$\chi^{2}$/NDF & NDF
\\
case & $[\%]$ &  & [MeV] & [MeV] & [fm] & [fm] &  & &
\\
\hline PHOBOS at & 0-15 & 294.0 & 148.52$\pm$1.15 & 72.60$\pm$3.81 &
7.84$\pm$0.19 & 8.35$\pm$0.14 & 0.684$\pm$0.006 & 0.995 & 72
\\
$\sqrt{s_{NN}}=62.4$ GeV & 15-30 & 175.0 & 149.64$\pm$1.42 &
69.47$\pm$4.13 & 6.23$\pm$0.19 & 7.20$\pm$0.17 & 0.654$\pm$0.007 &
0.43 & 72
\\
& 30-50 & 88.0 & 151.29$\pm$1.70 & 67.15$\pm$4.47 & 4.60$\pm$0.18 &
6.05$\pm$0.18 & 0.605$\pm$0.009 & 0.20 & 71
\\
\hline & 0-5 & 347.7 & 166.74$\pm$3.96 & 35.06$\pm$8.97 &
6.31$\pm$0.41 & 8.08$\pm$0.44 & 0.616$\pm$0.014 & 0.53 & 78
\\
PHENIX at & 5-15 & 271.3 & 161.70$\pm$3.21 & 43.14$\pm$7.77 &
6.34$\pm$0.35  & 7.57$\pm$0.34 & 0.642$\pm$0.012 & 0.46 & 78
\\
$\sqrt{s_{NN}}=130$ GeV & 15-30 & 180.2 & 162.33$\pm$3.29 &
38.52$\pm$7.75 & 5.32$\pm$0.29  & 6.54$\pm$0.29 & 0.631$\pm$0.012 &
0.50 & 78
\\
& 30-60 & 78.5 & 162.25$\pm$3.46 & 31.80$\pm$7.87 & 3.77$\pm$0.23 &
4.95$\pm$0.23 & 0.606$\pm$0.015 & 0.75 & 78
\\
& 60-92 & 14.3 & 159.46$\pm$6.85 & 37.05$\pm$16.09 & 1.87$\pm$0.27 &
3.26$\pm$0.27 & 0.496$\pm$0.043 & 1.37 & 42
\\
\hline & 0-5 & 351.4 & 150.07$\pm$1.34 & 24.10$\pm$3.66 &
9.28$\pm$0.21 & 9.48$\pm$0.19 & 0.699$\pm$0.004 & 0.69 & 122
\\
PHENIX at & 5-10 & 299.0 & 150.18$\pm$1.35 & 23.48$\pm$3.65 &
8.75$\pm$0.20  & 8.80$\pm$0.18 & 0.705$\pm$0.004 & 0.50 & 122
\\
$\sqrt{s_{NN}}=200$ GeV & 10-15 & 253.9 & 150.16$\pm$1.35 &
22.75$\pm$3.65 & 8.25$\pm$0.19  & 8.20$\pm$0.17 & 0.709$\pm$0.004 &
0.37 & 122
\\
& 15-20 & 215.3 & 150.00$\pm$1.36 & 22.38$\pm$3.65 & 7.80$\pm$0.18 &
7.69$\pm$0.16 & 0.712$\pm$0.004 & 0.37 & 122
\\
& 20-30 & 166.6 & 149.59$\pm$1.31 & 24.03$\pm$3.47 & 7.13$\pm$0.16 &
6.96$\pm$0.14 & 0.716$\pm$0.004 & 0.45 & 122
\\
& 30-40 & 114.2 & 149.79$\pm$1.36 & 23.78$\pm$3.56 & 6.14$\pm$0.14 &
6.03$\pm$0.12 & 0.713$\pm$0.004 & 0.66 & 122
\\
& 40-50 & 74.4 & 148.53$\pm$1.40 & 22.52$\pm$3.71 & 5.28$\pm$0.13 &
5.27$\pm$0.11 & 0.708$\pm$0.005 & 0.89 & 122
\\
& 50-60 & 45.5 & 147.75$\pm$1.51 & 22.02$\pm$4.03 & 4.38$\pm$0.12 &
4.55$\pm$0.10 & 0.694$\pm$0.005 & 0.96 & 122
\\
& 60-70 & 25.7 & 144.57$\pm$1.65 & 21.63$\pm$4.56 & 3.63$\pm$0.11 &
3.91$\pm$0.09 & 0.680$\pm$0.006 & 1.12 & 122
\\
& 70-80 & 13.4 & 141.77$\pm$1.98 & 24.13$\pm$5.68 & 2.84$\pm$0.10 &
3.22$\pm$0.09 & 0.662$\pm$0.008 & 1.23 & 122
\\
& 80-92 & 6.3 & 140.62$\pm$2.46 & 14.29$\pm$7.12 & 2.24$\pm$0.10 &
2.77$\pm$0.09 & 0.630$\pm$0.011 & 1.13 & 122
\\
\hline & 0-5 & 352.0 & 159.99$\pm$1.19 & 24.00$\pm$2.17 &
9.22$\pm$0.31 & 7.13$\pm$0.19 & 0.791$\pm$0.006 & 0.30 & 71
\\
STAR at & 5-10 & 299.0 & 160.58$\pm$1.16 & 24.97$\pm$2.17 &
8.34$\pm$0.28 & 6.75$\pm$0.18 & 0.777$\pm$0.006 & 0.27 & 71
\\
$\sqrt{s_{NN}}=200$ GeV & 10-20 & 234.0 & 161.20$\pm$1.14 &
22.91$\pm$2.15 & 7.45$\pm$0.24 & 6.17$\pm$0.16 & 0.770$\pm$0.006 &
0.22 & 73
\\
& 20-30 & 166.0 & 162.27$\pm$1.12 & 23.05$\pm$2.17 & 6.31$\pm$0.20 &
5.60$\pm$0.14 & 0.748$\pm$0.007 & 0.25 & 75
\\
& 30-40 & 115.0 & 161.97$\pm$1.08 & 20.43$\pm$2.17 & 5.38$\pm$0.17 &
5.15$\pm$0.12 & 0.722$\pm$0.007 & 0.19 & 75
\\
& 40-50 & 76.0 & 162.97$\pm$1.08 & 21.01$\pm$2.21 & 4.46$\pm$0.14 &
4.64$\pm$0.11 & 0.693$\pm$0.008 & 0.13 & 75
\\
& 50-60 & 47.0 & 163.41$\pm$1.07 & 18.75$\pm$2.25 & 3.67$\pm$0.12 &
4.13$\pm$0.10 & 0.664$\pm$0.008 & 0.13 & 75
\\
& 60-70 & 27.0 & 162.39$\pm$1.06 & 16.47$\pm$2.31 & 2.95$\pm$0.10 &
3.79$\pm$0.09 & 0.614$\pm$0.010 & 0.26 & 75
\\
& 70-80 & 14.0 & 163.70$\pm$1.15 & 15.84$\pm$2.50 & 2.22$\pm$0.09 &
3.16$\pm$0.08 & 0.574$\pm$0.012 & 0.61 & 75
\\
\hline & 0-10 & 328.0 & 150.60$\pm$1.39 & 23.07$\pm$3.51 &
9.26$\pm$0.25 & 8.65$\pm$0.21 & 0.731$\pm$0.004 & 0.43 & 114
\\
BRAHMS at & 10-20 & 239.0 & 151.38$\pm$1.48 & 26.53$\pm$3.72 &
8.07$\pm$0.23 & 7.68$\pm$0.19 & 0.724$\pm$0.005 & 0.42 & 114
\\
$\sqrt{s_{NN}}=200$ GeV & 20-40 & 140.0 & 149.43$\pm$1.54 &
25.92$\pm$3.98 & 7.00$\pm$0.21 & 6.73$\pm$0.17 & 0.721$\pm$0.005 &
0.26 & 112
\\
& 40-60 & 62.0 & 148.36$\pm$2.02 & 26.69$\pm$5.21 & 5.02$\pm$0.20 &
5.38$\pm$0.17 & 0.683$\pm$0.008 & 0.52 & 112
\\
\hline
\end{tabular}
\end{ruledtabular}
\end{table*}

\begin{figure}
\includegraphics[width=0.4\textwidth]{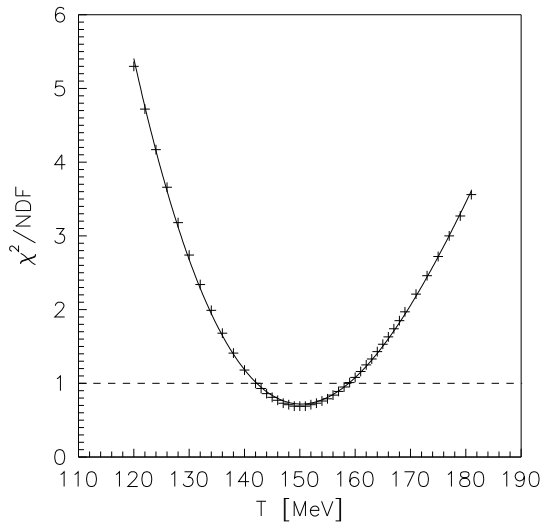}
\caption{\label{Fig.1} Simulation of the temperature dependence of
$\chi^{2}/$NDF for PHENIX at $\sqrt{s_{NN}}=200$ GeV and the $0-5
\%$ centrality class. Solid line is the polynomial 3 best fit.}
\end{figure}

The fitted results for $T_{kin}$, $\mu_{B}$, $\rho_{max}$ and $\tau$
are collected in Table~\ref{Table1} together with values of the
surface velocity $\beta_{\perp}^{max}$ and values of $\chi^{2}$/NDF
for each centrality class additionally characterized by the number
of participants $N_{part}$. Errors are expressed by one standard
deviation ($1\sigma$). Note that except the most peripheral bins of
the PHENIX measurements all fits are statistically significant. That
these new higher values of temperature (higher in comparison with
the blast-wave fit values) are at minima is shown in
Fig.~\ref{Fig.1} in an example of the temperature dependence of
$\chi^{2}$ for the PHENIX most central bin at $\sqrt{s_{NN}}=200$
GeV. The simulation of the dependence was done within the same
procedure as in the original single-freeze-out model, with the
overall fixed value of $\mu_{B}=27.5$ MeV (the average of values
from \cite{Rafelski:2004dp,Baran:2003nm}) and the temperature fixed
at each point and taken from the range 120-180 MeV with the 1 or 2
MeV step. Then two geometric parameters were fitted and $\chi^{2}$
at their optimum values was put into the figure.

\begin{figure}
\includegraphics[width=0.75\textwidth]{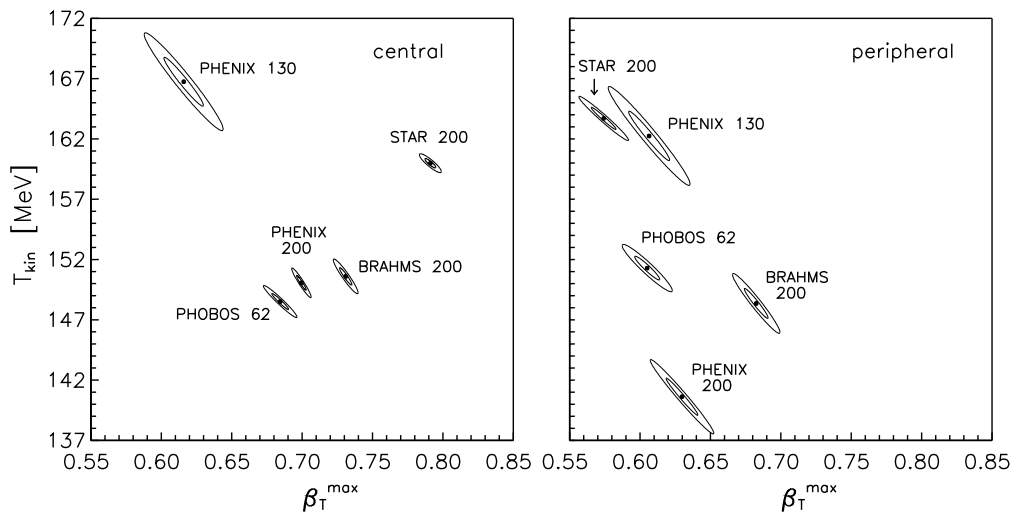}
\caption{\label{Fig.2} The $\chi^{2}$ contours ($1\sigma$ and
$2\sigma$) in the parameter plane $T_{kin}$ and
$\beta_{\perp}^{max}$ fixed by taking $\mu_{B}$ and $\tau$ at their
optimum values. In the right panel the PHENIX case at
$\sqrt{s_{NN}}=130$ GeV is represented by the second from the most
peripheral bin, see text for explanations. Dots denotes the optimum
values of $T_{kin}$ and $\beta_{\perp}^{max}$. }
\end{figure}

\begin{figure}
\includegraphics[width=0.75\textwidth]{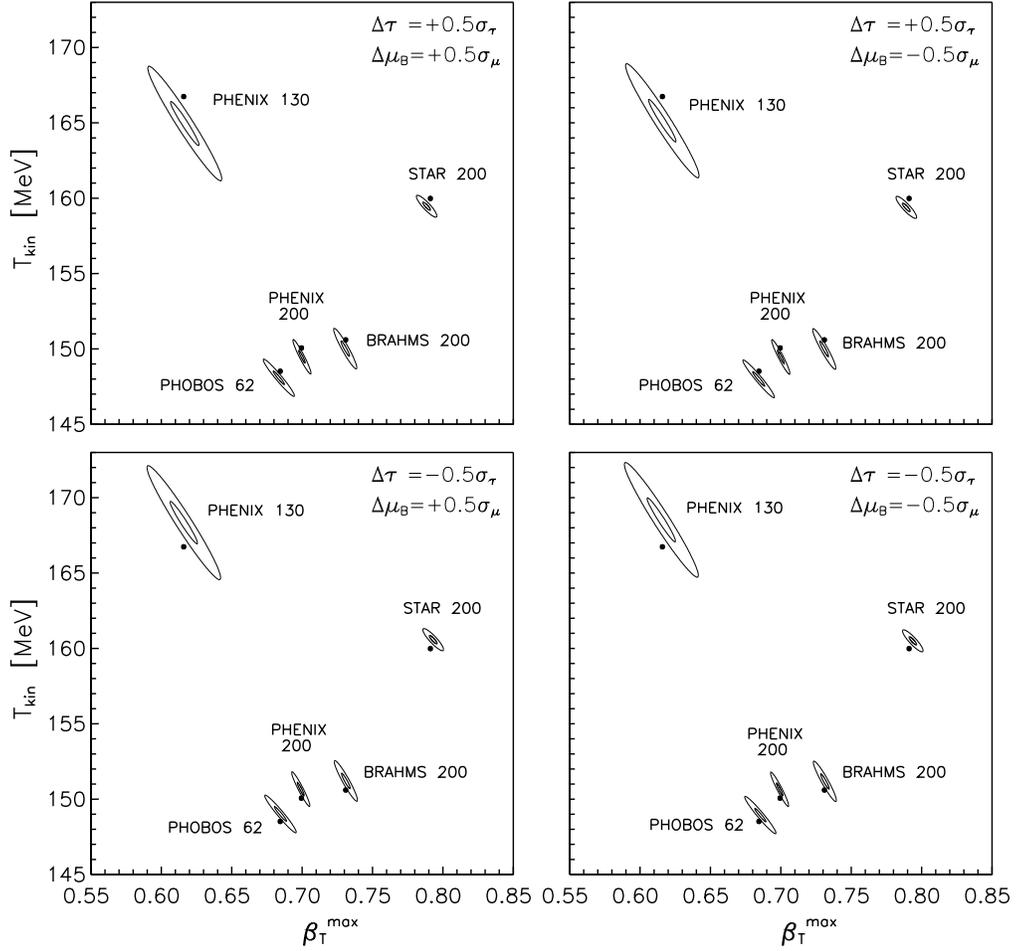}
\caption{\label{Fig.3} The $\chi^{2}$ contours ($1\sigma$ and
$2\sigma$) in the parameter planes $T_{kin}$ and
$\beta_{\perp}^{max}$ fixed by taking $\mu_{B}$ and $\tau$ at $\pm
0.5 \sigma$ from their optimum values, $\triangle \tau = \tau-
\tau_{opt}$, $\triangle \mu_{B}=\mu_{B}- \mu_{B,opt}$. All cases
represent the most central classes. Dots denote the projections of
the points of optimum values of all parameters onto the appropriate
plane $T_{kin}\;$-$\;\beta_{\perp}^{max}$. }
\end{figure}

Coming back to the results of simultaneous fits of all four
parameters, some $1\sigma$ and $2\sigma$ contours are presented in
Figs.~\ref{Fig.2} and \ref{Fig.3}. These contours are done in
$T_{kin}$ and $\beta_{\perp}^{max}$ planes to make them comparable
with the blast-wave results. But since the parameter space is four
dimensional, the $n$-sigma contour is in fact the 3-dimensional
ellipsoid embedded in this space, so it can be presented only by
means of its projections on some planes. Contours presented in
Figs.~\ref{Fig.2} and \ref{Fig.3} are such projections onto
$T_{kin}$ and $\beta_{\perp}^{max}$ planes chosen at some fixed
$\mu_{B}$ and $\tau$, namely at their optimum values $\mu_{B,opt}$
and $\tau_{opt}$, Fig.~\ref{Fig.2}, and at $\mu_{B}=\mu_{B,opt} \pm
0.5 \sigma_{\mu}$ and $\tau=\tau_{opt} \pm 0.5 \sigma_{\tau}$,
Fig.~\ref{Fig.3}. Note that in Fig.~\ref{Fig.3} black dots denote
the projections of the points of optimum values of all parameters
onto the appropriate plane $T_{kin}\;$-$\;\beta_{\perp}^{max}$, this
is the reason why they are outside the contours. In the right panel
of Fig.~\ref{Fig.2}, all examples are for most peripheral bins,
except the PHENIX case at $\sqrt{s_{NN}}=130$ GeV which is for the
second from the most peripheral class. This is because errors in the
case of the most peripheral bin of the PHENIX measurements at
$\sqrt{s_{NN}}=130$ GeV are substantially greater (see
Table~\ref{Table1}), which results in too big extension of $1\sigma$
and $2\sigma$ contours ($\sim 3$ times bigger than in the second
from the most peripheral bin). This fact together with
$\chi^{2}$/NDF much higher than 1 does not enable to determine firm
optimum values of $T_{kin}$ and $\beta_{\perp}^{max}$ in this case.
Generally, $\chi^{2}$ is flatter in the vicinity of the optimum
points for PHENIX at $\sqrt{s_{NN}}=130$ GeV, as one can notice from
Figs.~\ref{Fig.2} and \ref{Fig.3}. This is also expressed by $\sim
2$ times greater errors of the fitted values in this case. For other
experiments $1\sigma$ and $2\sigma$ contours are relatively small.
Their sizes do not change visibly from most central to mid-central
bins and then increase gradually up to the sizes of the contours for
most peripheral bins (in fact this behavior reflects the behavior of
the corresponding errors, cf. Table~\ref{Table1}).

The feeding from weak decays is treated in the same way as
experimental groups do, except the STAR case. So, for the PHENIX
measurements at $\sqrt{s_{NN}}=200$ GeV protons (antiprotons) from
$\Lambda$ ($\bar{\Lambda}$) decays are excluded. In the PHOBOS case
protons (antiprotons) from $\Lambda$ and $\Sigma^{+}$
($\bar{\Lambda}$ and $\bar{\Sigma}^{-}$) decays are not counted. As
far as STAR is considered, this Collaboration has corrected its pion
spectra for weak decays \cite{Adams:2003xp}. However, STAR also
claims there that when the full feeding from weak decays is taken
into account the single-freeze-out model with all the resonances can
fit the spectra but with higher $\chi^{2}$/NDF. Therefore, to check
this statement all weak decays are included in the STAR data
analysis.

\begin{figure}
\includegraphics[width=0.4\textwidth]{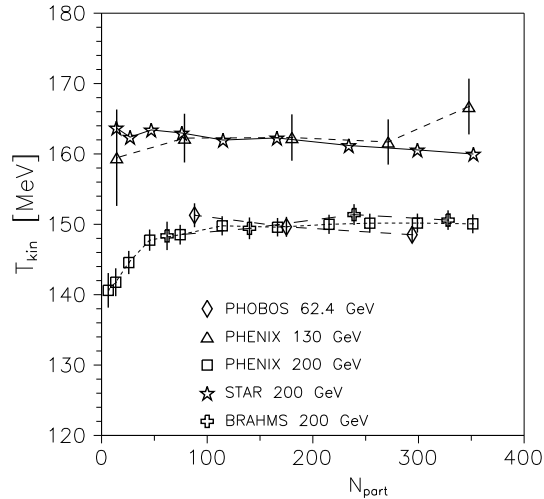}
\caption{\label{Fig.4} Centrality dependence of the kinetic
freeze-out temperature for the RHIC measurements at
$\sqrt{s_{NN}}=62.4,\; 130$ and $200$ GeV. The lines connect the
results and are a guide. }
\end{figure}

\begin{figure}
\includegraphics[width=0.4\textwidth]{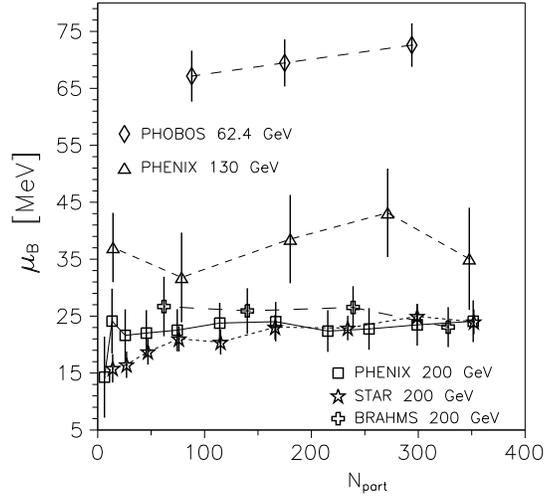}
\caption{\label{Fig.5} Centrality dependence of the baryon number
chemical potential for the RHIC measurements at
$\sqrt{s_{NN}}=62.4,\; 130$ and $200$ GeV. The lines connect the
results and are a guide. }
\end{figure}

Results for $T_{kin}$ and $\mu_{B}$ are also depicted as functions
of $N_{part}$ in Figs.~\ref{Fig.4} and \ref{Fig.5}, respectively. It
is clearly seen that both $T_{kin}$ and $\mu_{B}$ are almost
independent of the collision centrality, only for peripheral bins
some dependence can be observed. Additionally, their values are very
close to the values at the chemical freeze-out. Namely,
$T_{chem}=165-169$ MeV and $\mu_{B}=33-38$ MeV from the peripheral
to most central bin at $\sqrt{s_{NN}}=130$ GeV was found in
Ref.~\cite{Cleymans:2004pp}, $T_{chem} \approx 155$ MeV and $\mu_{B}
\approx 26$ MeV independent of the centrality for PHENIX at
$\sqrt{s_{NN}}=200$ GeV in Ref.~\cite{Rafelski:2004dp} and $T_{chem}
\approx 160$ MeV independent of the centrality and $\mu_{B} =15-24$
MeV from the peripheral to most central bin for STAR at
$\sqrt{s_{NN}}=200$ GeV in
Refs.~\cite{Adams:2003xp,Barannikova:2004rp,Barannikova:2005rw}.

\begin{figure}
\includegraphics[width=1.0\textwidth]{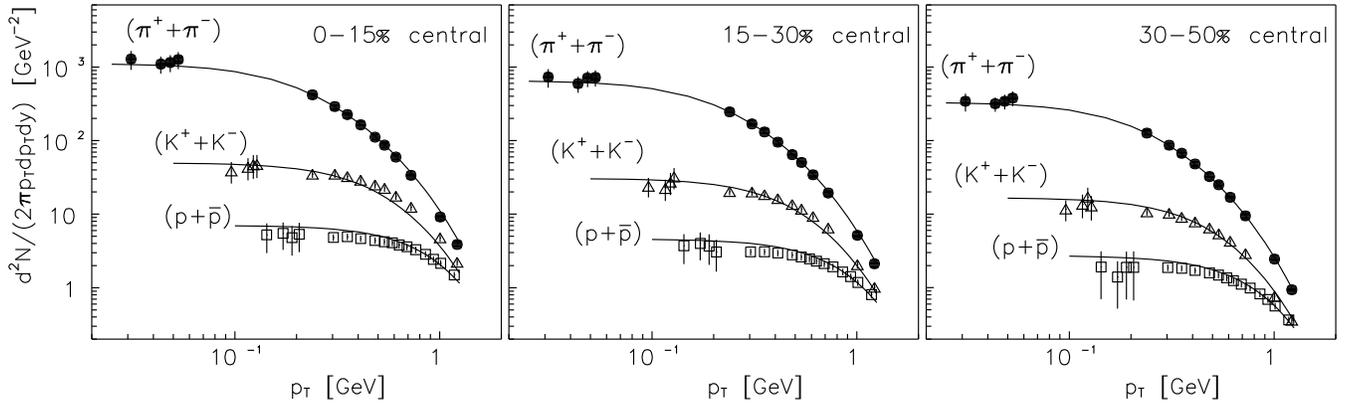}
\caption{\label{Fig.6} Invariant yields as a function of $p_{T}$ in
Au+Au collisions at $\sqrt{s_{NN}}=62.4$ GeV. Lines are predictions
of the present model and symbols are PHOBOS data
\protect\cite{Back:2006tt}. }
\end{figure}

\begin{figure}
\includegraphics[width=0.5\textwidth]{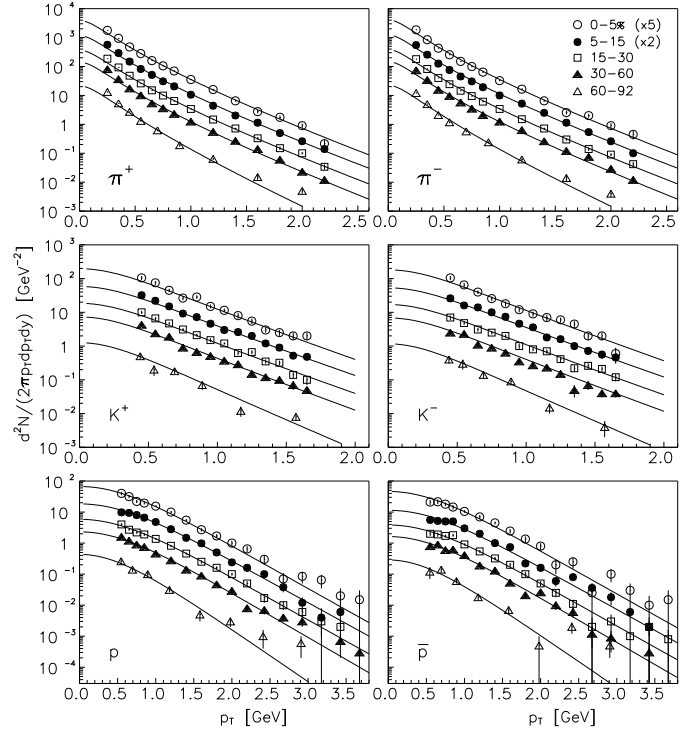}
\caption{\label{Fig.7} Invariant yields as a function of $p_{T}$ in
Au+Au collisions at $\sqrt{s_{NN}}=130$ GeV. Lines are predictions
of the present model and symbols are PHENIX data
\protect\cite{Adcox:2003nr}. For clarity, the data points are scaled
vertically for two bins as quoted in the figure.}
\end{figure}

\begin{figure}
\includegraphics[width=0.8\textwidth]{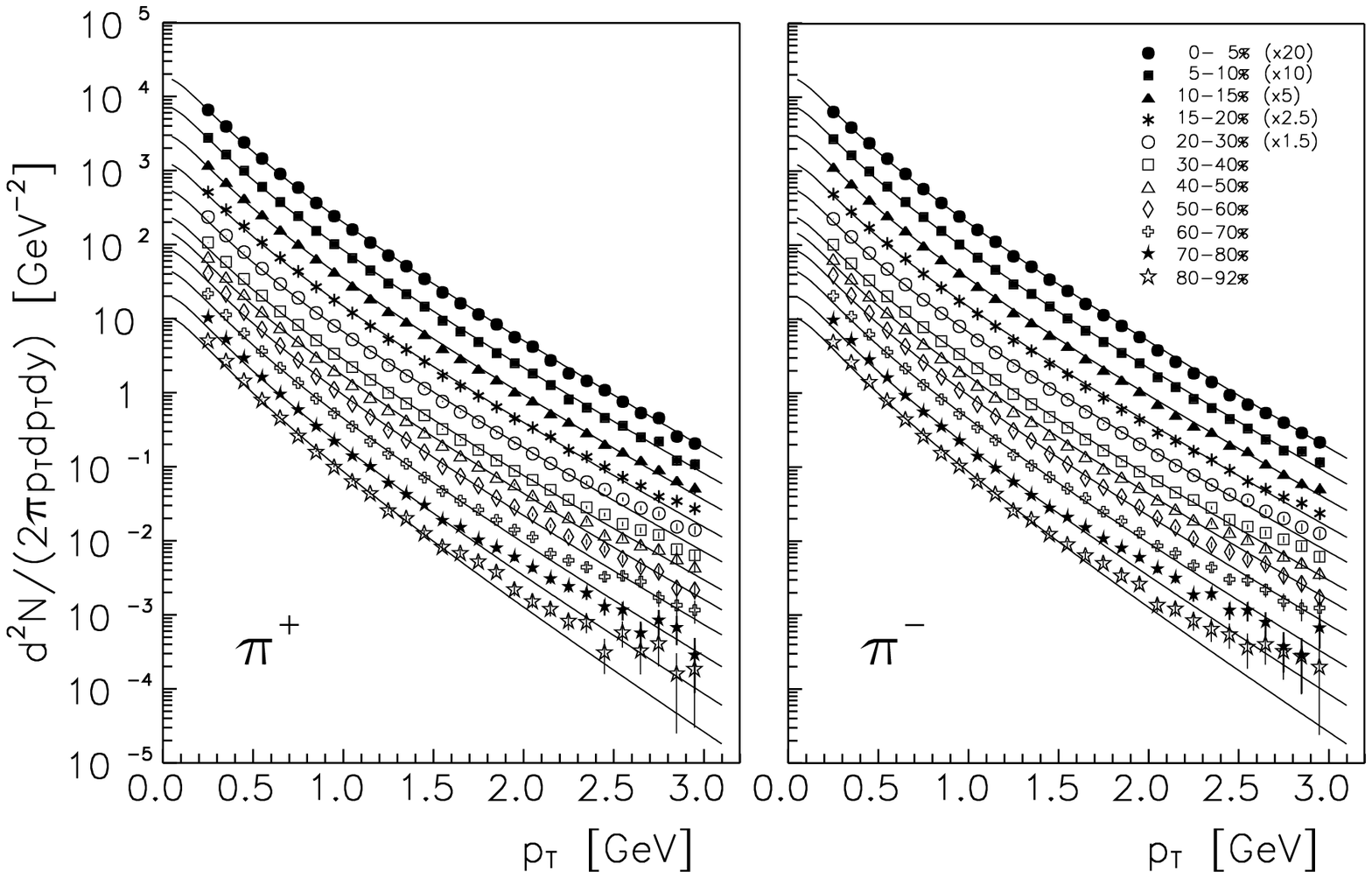}
\caption{\label{Fig.8} Invariant yields of $\pi^{+}$ (left) and
$\pi^{-}$ (right) as a function of $p_{T}$ in Au+Au collisions at
$\sqrt{s_{NN}}=200$ GeV. Lines are predictions of the present model
and symbols are PHENIX data \protect\cite{Adler:2003cb}. For
clarity, the data points are scaled vertically for five bins as
quoted in the figure. }
\end{figure}

\begin{figure}
\includegraphics[width=0.8\textwidth]{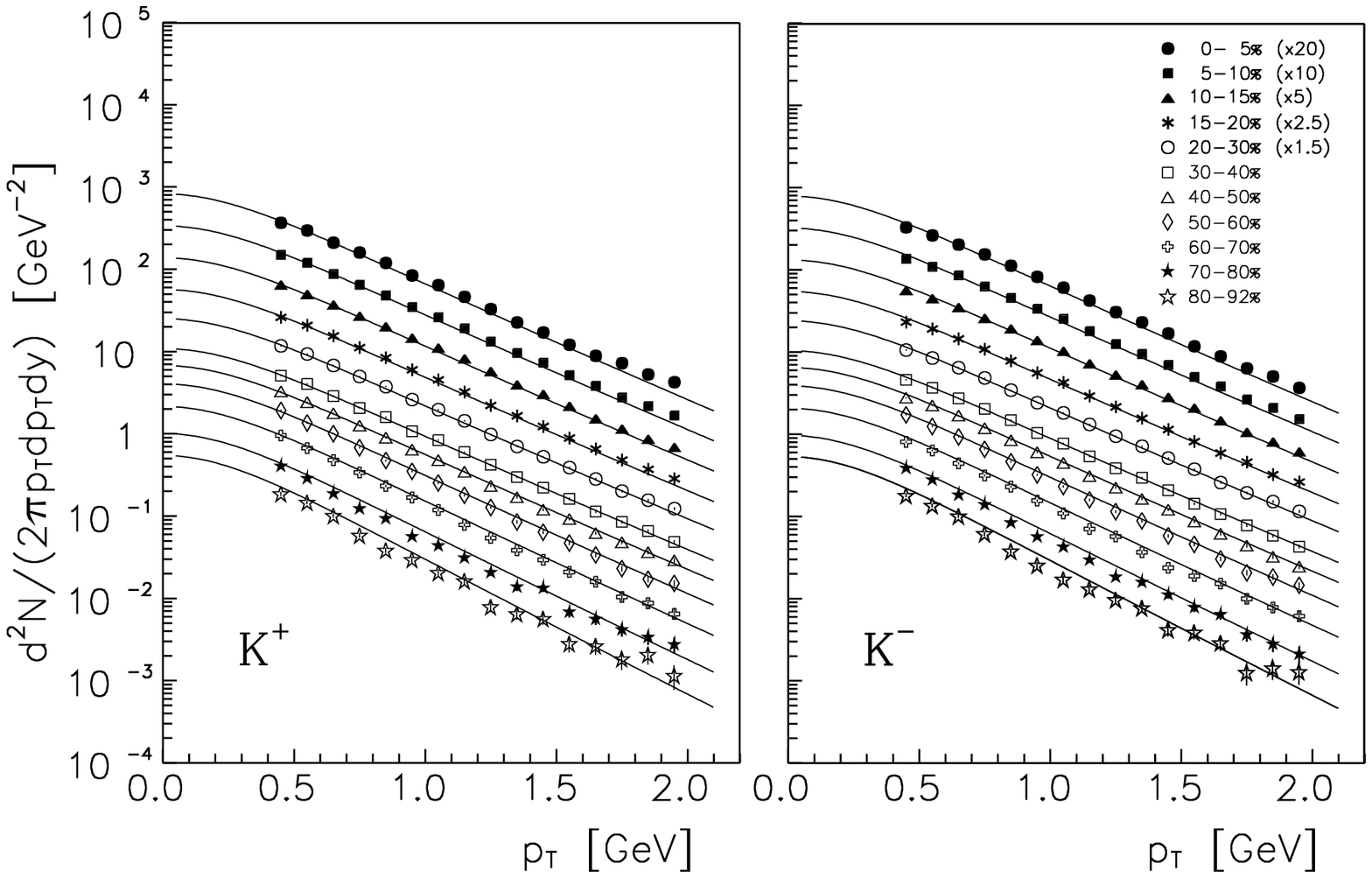}
\caption{\label{Fig.9} Invariant yields of $K^{+}$ (left) and
$K^{-}$ (right) as a function of $p_{T}$ in Au+Au collisions at
$\sqrt{s_{NN}}=200$ GeV. Lines are predictions of the present model
and symbols are PHENIX data \protect\cite{Adler:2003cb}. For
clarity, the data points are scaled vertically for five bins as
quoted in the figure. }
\end{figure}

\begin{figure}
\includegraphics[width=0.8\textwidth]{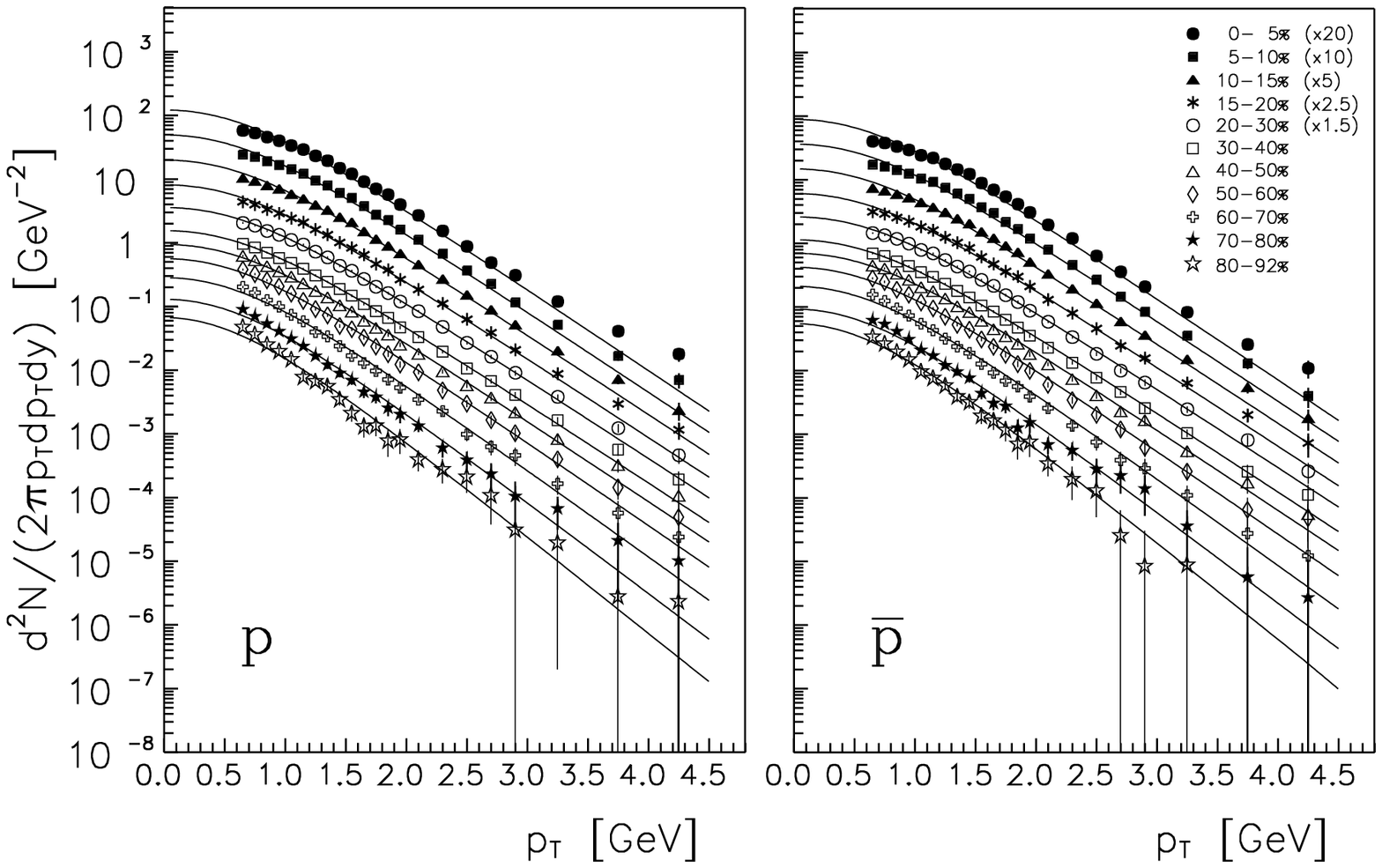}
\caption{\label{Fig.10} Invariant yields of protons (left) and
anti-protons (right) as a function of $p_{T}$ in Au+Au collisions at
$\sqrt{s_{NN}}=200$ GeV. Lines are predictions of the present model
and symbols are PHENIX data \protect\cite{Adler:2003cb}. For
clarity, the data points are scaled vertically for five bins as
quoted in the figure. }
\end{figure}

In Figs.~\ref{Fig.6}-\ref{Fig.10} measured and predicted spectra of
identified hadrons are presented for some collisions at RHIC. In
Fig.~\ref{Fig.6} the case of PHOBOS at $\sqrt{s_{NN}}=62.4$ GeV is
dealt with. The way the data and results are depicted is suggested
by the PHOBOS analysis \cite{Back:2006tt} (see Fig. 7 there). The
lines are predictions of the present model done with the use of
fitted parameters tabulated in Table~\ref{Table1}. But fits were
done within intermediate ranges of $p_{T}$, that is in the ranges
where single charged hadron data exist. Then predictions were made
for sums of negatively and positively charged hadrons of the same
kind but in the whole accessible ranges, namely ranges extended to
the very low $p_{T}$ where separate data on the summed yields exist.
As one can see from Fig.~\ref{Fig.6} predictions of the model agree
very well with the data, only slight overestimation of protons and
antiprotons can be observed but for the very low $p_{T}$ region
results agree within errors. Predictions of the blast-wave model
also agree with the data (cf. Fig. 7 in \cite{Back:2006tt}), but
opposite to the present analysis, they underestimate pion yields and
correctly estimate proton and antiproton yields at very low $p_{T}$.
For the collisions at $\sqrt{s_{NN}}=130$ GeV, Fig.~\ref{Fig.7},
predictions for pion and kaon spectra agree very well with the
PHENIX data, whereas for protons and antiprotons the agreement holds
up to $p_{T} \approx 2.5$ GeV. Similar results were obtained within
the blast-wave model (cf. Fig. 19 in \cite{Adcox:2003nr}). In
Figs.~\ref{Fig.8}-\ref{Fig.10} results and data are presented for
the PHENIX measurements at $\sqrt{s_{NN}}=200$ GeV. Predictions for
pions, Fig.~\ref{Fig.8}, agree very well with the data practically
in the whole $p_{T}$ range for the first five bins from the top, for
the next bins the predictions start to miss the data at $p_{T}
\approx 2.5$ GeV and this value is decreasing with the centrality to
$p_{T} \approx 2$ GeV for the most peripheral bin. Kaons are
reproduced very well in the whole range of $p_{T}$ and for all
centralities, as one can see in Fig.~\ref{Fig.9}. Protons and
antiprotons agree very well up to $p_{T} \approx 3.3$ GeV for first
nine bins from the top, for the two last peripheral bins the
agreement is lost at $p_{T} \approx 2.5$ GeV, as it can be seen in
Fig.~\ref{Fig.10}.

\begin{figure}
\includegraphics[width=0.4\textwidth]{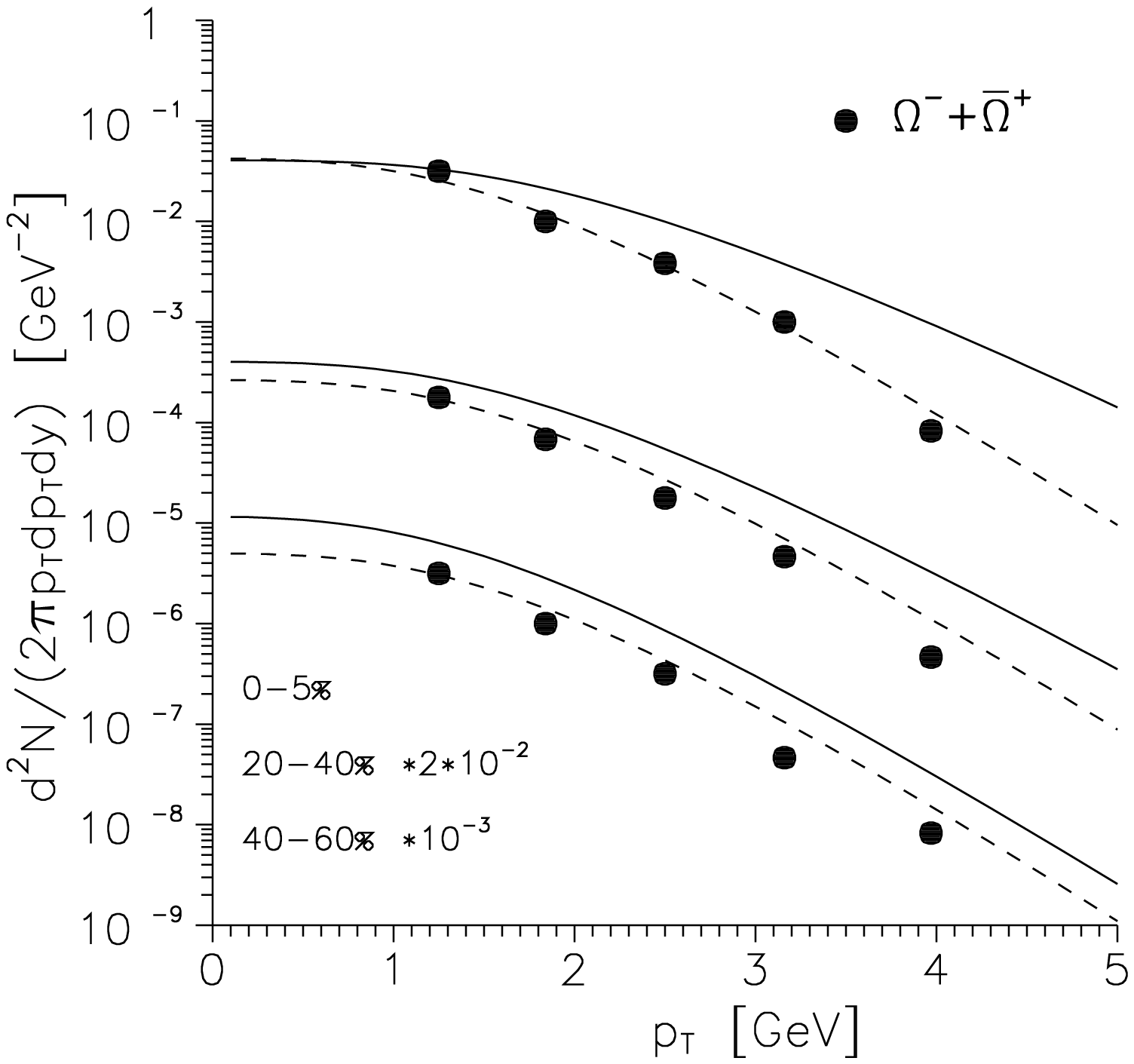}
\caption{\label{Fig.11} Transverse momentum distributions of
$\Omega^{-}+\bar{\Omega}^{+}$ for $\mid y \mid < 0.75$ in Au-Au
collisions at $\sqrt{s_{NN}}=200$ GeV. Data are from
\cite{Adams:2006ke} (STAR) scaled for clarity, (statistical) errors
are of the size of symbols. Lines denote model predictions: solid
based on fits to the STAR spectra, dashed based on fits to the
PHENIX spectra. }
\end{figure}

\begin{figure}
\includegraphics[width=0.75\textwidth]{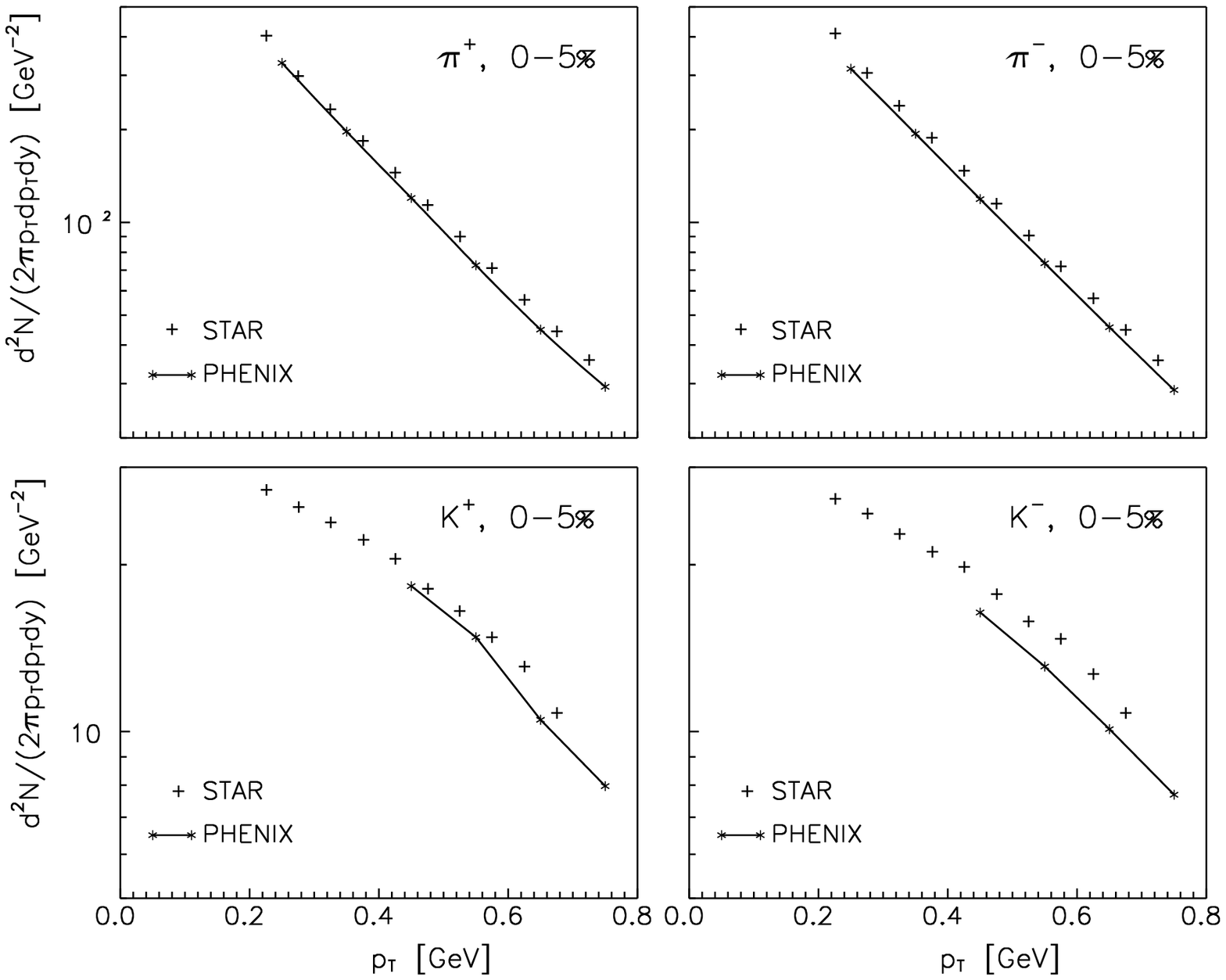}
\caption{\label{Fig.12} Comparison of $\pi^{+}$, $\pi^{-}$, $K^{+}$
and $K^{-}$ spectra measured by STAR and PHENIX for the $0-5 \%$
centrality bin at $\sqrt{s_{NN}}=200$ GeV. All STAR data points are
depicted, whereas PHENIX data ranges are cut from the right side (in
fact they extend to $p_{T}=2.95$ GeV for pions and $p_{T}=1.95$ GeV
for kaons). }
\end{figure}

A simple test of the model can be performed with the use of the
spectra of $\Omega$ hyperon. This is because $\Omega$ has only the
thermal contribution to the invariant distribution,
eq.~(\ref{Cooper2}). Such a test has been done for the blast-wave
model \cite{Barannikova:2004rp} and for the single-freeze-out model
\cite{Baran:2003nm}, but in both cases the comparison was done with
preliminary data. Present results together with the STAR data for
$\Omega^{-}+\bar{\Omega}^{+}$ production at $\sqrt{s_{NN}}=200$ GeV
are shown in Fig.~\ref{Fig.11}. Values of parameters for $20-40 \%$
and $40-60 \%$ centrality bins explored by STAR in $\Omega$
measurements are the averages of the values from Table~\ref{Table1}
for bins which added percent coverage equals $20-40 \%$ and $40-60
\%$, respectively. One can see that predictions based on fits to
PHENIX spectra agree well with the data. Predictions based on fits
to STAR spectra agree only qualitatively, they have higher
normalization. Also for the $0-5 \%$ bin the slope differs.
Blast-wave model predictions for $\Omega^{-}+\bar{\Omega}^{+}$
spectrum for the $0-5 \%$ bin of the preliminary STAR data were done
in Ref.~\cite{Barannikova:2004rp}, but they do not agree with the
data neither in normalization nor in a slope. The probable reason
for the worse agreement of the STAR data based predictions in the
present model, Fig.~\ref{Fig.11}, is that STAR spectra of identified
stable hadrons are measured in narrower ranges of $p_{T}$ than
PHENIX ones, i.e. $p_{T} \in [0.2, 1.2]$ GeV/c approximately for
STAR \cite{Adams:2003xp} whereas for PHENIX $p_{T} \in [0.25, 2.95]$
GeV/c (pions), $p_{T} \in [0.55, 1.95]$ GeV/c (kaons) and $p_{T} \in
[0.65, 4.25]$ GeV/c [(anti)protons] \cite{Adler:2003cb}. And the
STAR measurement of $\Omega^{-}+\bar{\Omega}^{+}$ is done within the
range $p_{T} \in [1.25, 4]$ GeV/c, practically outside the STAR
range of $p_{T}$ of identified stable hadrons but covering the great
part of PHENIX $p_{T}$ ranges. Also they differ in common ranges of
$p_{T}$, namely STAR spectra are placed slightly above the
corresponding PHENIX spectra and in the case of pions have different
slopes (it has been checked carefully for the common $0-5 \%$ bin
after conversion of STAR spectra from $m_{T}-m_{i}$ to $p_{T}$, see
Fig.~\ref{Fig.12}). Actually, the difference in pion spectra is even
greater than what can be seen in Fig.~\ref{Fig.12}, because the STAR
pion spectra are corrected for weak decays \cite{Adams:2003xp}.
Proton and antiproton spectra are not compared, since the PHENIX
subtracted protons (antiprotons) from $\Lambda$ ($\bar{\Lambda}$)
decays, so by definition the STAR proton (antiproton) spectra are
above the PHENIX ones. Also that the STAR minimizations have lower
$\chi^{2}$/NDF than PHENIX ones is probably because it is much
easier to fit over a narrower range of $p_{T}$. Discrepancies
between STAR and PHENIX spectra explain why the optimum values of
parameters are different for these collaborations. These
discrepancies could be caused by the different detectors used by the
aforementioned Collaborations. The PHENIX uses a spectrometer, which
consists of drift chamber (DC), pad chamber (PC) and time-of-flight
(TOF) \cite{Adler:2003cb}, whereas the STAR uses time projection
chamber (TPC) \cite{Adams:2003xp}. The TPC is better in measurements
of resonances but the PHENIX spectrometer measures spectra of
identified particles more precisely
\cite{Rafelski:2004dp,Rafelski:2006nv}. This might also help to
understand why $\Omega^{-}+\bar{\Omega}^{+}$ measured spectra (an
example of the STAR data on resonance production) are well
reproduced in the model with the use of the parameters fitted to the
PHENIX data on identified stable hadrons (in addition to the earlier
argument based on the compatibility of $p_{T}$ ranges).

\section {Discussion and conclusions}
\label{Conclus}

Before the final conclusion one result of ref.~\cite{Heinz:2006ur}
should be commented. In \cite{Heinz:2006ur} a hydrodynamical model
supplemented with the dynamical freeze-out criterium
\cite{Bondorf:1978kz,Lee:1987ku,Bertsch:1987ux}

\begin{equation}
{1 \over \tau_{scatt}} = \xi \; {1 \over \tau_{exp}} = \xi \;
\partial_{\mu}u^{\mu}, \label{Rates}
\end{equation}

\noindent where ${1/\tau_{scatt}}$ is the local scattering rate and
${1/\tau_{exp}}$ is the local expansion rate and the parameter $\xi
\sim 1$, was successfully used to explain the centrality dependence
and the temperature range of $T_{kin}$ determined from the
blast-wave fits to the STAR data at $\sqrt{s_{NN}}=200$ GeV
\cite{Adams:2003xp}. To obtain the $T_{kin}$ range comparable with
the STAR results, $T_{kin}\simeq 90-130$ MeV, the appropriate
adjustment of the parameter $\xi$ was done, $\xi=0.295$. Then, the
range $T_{kin}\simeq 105-135$ MeV was obtained in
\cite{Heinz:2006ur}. According to the author's knowledge the
parameter $\xi$ appeared the first time in \cite{Heinz:2006ur} just
to fix the range of temperature as was explained above.

Generally, the dynamical criterium for freeze-out was formulated as
the moment when the local velocity of the rarefaction due to the
expansion overcomes the local thermal velocity
\cite{Bondorf:1978kz}. Since the freeze-out means the end of
collisions the criterium can be stated more precisely as the moment
at a space point when the average local proper time between
subsequent collisions $\tau_{scatt}$ overcomes the characteristic
time for expansion of the system given by $\tau_{exp} = V/\dot{V}$,
where $\dot{V}$ denotes the derivative of the volume of the system
with respect to the local proper time. The expansion characteristic
time is the time which the evolution of the system from very small
volume (almost zero) to the given volume $V$ would take with the
constant speed equal to the local velocity $\dot{V}$. For a given
hadron species the scattering time can be estimated as $\tau_{scatt}
= \lambda/\upsilon_{th} = 1/\rho_{tot}\sigma_{tot}\upsilon_{th}$,
where $\lambda$ is the mean free path, $\rho_{tot}$ is the total
density of all particles, $\sigma_{tot}$ is the average cross
section and $\upsilon_{th}$ is the average thermal velocity
\cite{Lee:1987ku,Bertsch:1987ux}. The characteristic expansion time
can be expressed as $\tau_{exp} = (\partial_{\mu}u^{\mu})^{-1} =
V/\dot{V}$ \cite{Eskola:2007zc}.

Hydrodynamics can be applied for $\tau_{scatt} \ll \tau_{exp}$ but
when $\tau_{scatt} \gg \tau_{exp}$ certainly a system is not in
statistical and thermal equilibrium \cite{Hung:1997du}. However the
determination of the strict border between a flow and a free stream
of particles is somehow arbitrary, it is commonly accepted that on
this border $\tau_{scatt} \sim \tau_{exp}$. Both times increase when
the system cools down, but $\tau_{scatt}$ is growing steeper, so it
cuts $\tau_{exp}$ at some temperature (it represents the case
$\xi=1$; for some examples of simulations of $\tau_{scatt}$ and
$\tau_{exp}$ see ref.~\cite{Bertsch:1987ux}). In terms of rates
$\tau_{scatt}^{-1}$ is falling steeper and cuts $\tau_{exp}^{-1}$ at
some temperature. For the expansion rate multiplied by $\xi$, if
$\xi < 1$ then $\xi \tau_{exp}^{-1}$ cuts $\tau_{scatt}^{-1}$ at
some lower temperature, if $\xi > 1$ it happens at some higher
temperature. Therefore, the choice $\xi=0.295$ as in
\cite{Heinz:2006ur} means decreasing the freeze-out temperature so
as it falls into the blast-wave range. Thus it is very likely that
increasing $\xi$ but still keeping $\xi \sim 1$ one could obtain the
freeze-out temperature in the range $T_{kin}\simeq 140-165$ MeV as
in the present model. Another point is the centrality dependence of
$T_{kin}$. The change of $T_{kin}$ with the centrality, of the order
of $25 \%$ between the most central and the most peripheral bins,
was obtained from eq.~(\ref{Rates}) with the same expression for
$\tau_{scatt}^{-1}$ taken for all centrality classes
\cite{Heinz:2006ur}. Such assumption might not be reasonable, since
e.g. in Ref.~\cite{Schnedermann:1994gc} variation of the temperature
dependence of $\tau_{scatt}$ with specific entropy $S/A= s/n_{B}$
(entropy density per baryon number density) was shown, see Fig. 7
there. The specific entropy, as an initial condition, could be
different for each centrality class, so the different expression for
$\tau_{scatt}^{-1}$ should be put into eq.~(\ref{Rates}) in each
case. Moreover, the scattering rate was calculated for pions only in
a pion-kaon mixture \cite{Hung:1997du}. The full scattering rate
should include interactions between all constituents of the gas. Of
course, this is an open question how such realistic scattering rate
could influence the centrality dependence of $T_{kin}$  determined
from eq.~(\ref{Rates}), but that this might flatten the dependence
can not be excluded. And the last remark, $T_{kin}$ points described
as the STAR data \cite{Adams:2003xp} in Fig. 3 of
Ref.~\cite{Heinz:2006ur} are not the measured quantities. They are
blast-wave-fit results. Thus Fig. 3 in \cite{Heinz:2006ur}
demonstrates that $T_{kin}$ obtained within the model where the
freeze-out takes place on a hypersurface of constant temperature is
consistent with the average $T_{kin}$ for the hypersurface on where
the temperature is not constant.

In summary, the alternative hydrodynamical model has been proposed
to describe hadronic $p_{T}$ spectra measured at relativistic
heavy-ion collisions. So far, conclusions about chemical and thermal
(kinetic) freeze-outs have been drawn from the blast-wave
parametrization \cite{Schnedermann:1993ws} of the final stage of the
collision . It has turned out that those conclusions are not
definite and depend strongly on the applied model. In the present
model the temperature and the baryon number chemical potential
fitted to the spectra are almost independent of the centrality of
the collision and their values are very close to the values at the
chemical freeze-out, what is opposite to the conclusions drawn from
the blast-wave model analysis \cite{Heinz:2006ur}. Such behavior
justifies the \textit{ad hoc} assumption about one freeze-out
postulated in Refs.~\cite{Broniowski:2001we,Broniowski:2001uk}.

\begin{acknowledgments}
This work was supported in part by the Polish Ministry of Science
and Higher Education under contract No. N N202 0953 33.
\end{acknowledgments}

\end{document}